# Intercalation and desorption of oxygen between graphene and Ru(0001) studied with helium ion scattering

Tianbai Li and Jory A. Yarmoff*

*Department of Physics and Astronomy, University of California, Riverside, Riverside CA 92521*

**Abstract**

Graphene is a fascinating 2D material that is being widely investigated for use in electronic devices due to its unique electronic and materials properties. Also, because of its high thermal stability and inertness, it is considered a promising candidate for use as a protection layer for metal substrates. Here, graphene films grown on Ru(0001) are held at 600 K while reacted with oxygen ($O_2$) and then investigated with helium low energy ion scattering (LEIS). LEIS spectra collected at different scattering angles confirm that oxygen does not adsorb to graphene, but instead intercalates between the graphene and the substrate. The intercalated $O_2$ desorbs when the sample is annealed to 800 K. It is shown that this is a much lower temperature than is needed to remove chemisorbed atomic oxygen from Ru, thus inferring that the intercalated oxygen is molecular. During the desorption process, some of the graphene is etched away via a chemical reaction with the oxygen, with the proportion desorbing as $O_2$ or reacting to etch the graphene being dependent on the amount of intercalated $O_2$.

___________

*Corresponding author, E-mail: yarmoff@ucr.edu

**I. Introduction**

Graphene (Gr) has attracted interest in many fields due to its special properties as a single $sp^2$ carbon layer [1-3]. Among them, its chemical and thermal stability suggest its use as a protection layer, especially on transition metal surfaces, to avoid oxidation or corrosion when working in an ambient environment [4,5]. Moreover, graphene is highly regarded as a promising candidate for carbon-based electronic devices [6]. During the preparation of devices with suspended graphene, the material is typically pre-annealed in vacuum to obtain better quality materials and performance [7]. Presumably, this annealing removes adsorbed contaminant species that modify the Gr electronic structure. Recent studies show, however, that small molecules, such as $O_2$, CO and $H_2O$, intercalate between a graphene overlayer and a metal substrate, rather than adsorb on the surface [8-11]. The contrasting claims of adsorption vs. intercalation of contaminants provide the motivation for studying the reaction between $O_2$ and Gr grown on a metal substrate, in this case Ru (0001), to see where the $O_2$ sticks and how oxygen exposure and removal affects the quality of the graphene.

The present study uses helium (He) low energy ion scattering (LEIS), which is well known for its extremely high surface sensitivity [12] and has been previously applied to measure impurities in Gr overlayers [13]. LEIS directly reveals the elemental composition and structural parameters of the outermost atomic layers. Helium ions experience Auger neutralization (AN) during scattering, which leads to a high neutralization probability that is proportional to time spent interacting with the surface leading to a strong surface sensitivity when detecting only the scattered ions. In addition, it is possible to probe either the $1^{st}$ or the $1^{st}$ and $2^{nd}$ atomic layers by adjusting the scattering angle, which provides a useful tool to explore the reaction of $O_2$ with Gr/Ru(0001). The LEIS results physically confirm that the $O_2$ intercalates between the Gr and the Ru substrate at 600 K and does not adsorb on top of the Gr, in agreement with Refs. [9,14]. Note that oxygen does not react with Gr/Ru(0001) at room temperature. It is also found that oxygen thermally desorbs from the sample at a relatively low temperature, suggesting that it remains as a molecule when intercalated. Furthermore, it is demonstrated that some of the Gr is etched during the desorption process.

## II. Experimental procedure

The experiments are carried out in an ultra-high vacuum (UHV) chamber with a base pressure of $4\times10^{-10}$ Torr. The chamber is equipped with an ion sputter gun (Varian) for sample cleaning, low energy electron diffraction (LEED) optics (Varian) and the instrumentation needed for LEIS.

The sample is mounted on a x-y-z manipulator that enables rotation about both the polar and azimuthal angles. The sample holder (VG) contains an e-beam heater that uses a rhenium-coated tungsten filament held at -650 V with respect to the sample. The temperature of the sample is measured by type K thermocouples attached to the sample plate.

The Ru(0001) substrate is cleaned using a standard ion bombardment and annealing (IBA) approach, as reported in the literature [14,15]. This involves 500 eV $Ar^+$ ion sputtering for 30 min with a spot size of $3\times3$ $cm^2$ and a flux of $4\times10^{13}$ ions/sec·$cm^2$, annealing under $4\times10^{-8}$ Torr of $O_2$ at 1100 K for 8 min followed by a flash annealing at 1300 K for 2 min under UHV. The reaction with $O_2$ is used to remove carbon contamination from the Ru substrate. Normally, this process is repeated several times to acquire a clean and well-ordered surface. The quality of the clean surface is verified with LEIS and LEED.

The graphene layer is prepared by a chemical vapor deposition (CVD) method in which the Ru surface is heated to 900 K and then exposed to $1.5\times10^{-7}$ Torr of ethylene for 5 min, followed by annealing under vacuum at 1200 K for 1 min and then slowly cooling down to 450 K for another 5 min [14]. This growth cycle is repeated several times until the surface is fully covered by a single, continuous graphene layer. The presence and quality of the Gr overlayer is confirmed by LEIS and LEED.

The $O_2$ exposures of the Gr/Ru(0001) samples are performed at a pressure of $1.5\times10^{-6}$ Torr with the sample held at 600 K. Exposures are reported in units of Langmiurs (L), where 1 L = $1\times10^{-6}$ torr sec. The sample is cooled to room temperature before the $O_2$ is evacuated from the chamber.

Helium ions for the LEIS measurements are generated by a differentially pumped ion gun (PHI model 04-303). The 3 keV incident ion beam has a diameter of 1.6 mm, and the current measured on the sample is 1.5 nA. The scattered ions are collected by a Comstock AC-901 160° hemispherical electrostatic analyzer (ESA) that has a radius of 47.6 mm. The ESA is mounted on a rotatable platform inside the chamber that allows the scattering angle to be adjusted. For the data collected here, a specular geometry is always used such that the incident and exit angles are equal with respect to the surface normal as the scattering angle is varied.

Note that ion scattering is an inherently destructive process, so it is important to verify that beam damage does not affect the results. To test whether beam damage plays a role in the present measurements, five spectra were collected successively from the same spot of a sample. Note that the sample was exposed to small amounts of $O_2$ between spectra, but this does not affect the size of the C peak because the oxygen is positioned below the Gr, as discussed below. It was found that the intensity of the carbon peak did not decrease during the test, which verifies that the amount of beam damage that occurs during the collection of 5 spectra is negligible. In the measurements reported here, the surface is re-prepared after every 4 spectra are collected to negate any effects of beam damage.

**III. Results**

For the 3 keV low energy ions used in these experiments, the de Broglie wavelength is so small that their wave properties can be neglected and the scattering process can be analyzed with classical mechanics. In addition, because the scattering cross sections are smaller than the interatomic spacings, it can be assumed the projectile interacts with only one target atom at a time, which is known as the binary collision approximation (BCA) [16]. In low energy ion scattering energy spectra, the most significant features are the single scattering peaks (SSPs) which correspond to projectiles that experience one hard, elastic collision with a target atom that causes the projectile to backscatter directly into the detector. The position of a SSP is determined primarily by the projectile/target mass ratio and the scattering angle, although continuous inelastic losses due to interaction with substrate electrons can slightly reduce the energy of the scattered projectile [12]. Since carbon and oxygen are both light elements, a very light projectile, such as helium, is needed to easily observe single scattering from these species.

The high probability for neutralization of scattered He ions is the main reason for the surface sensitivity of He LEIS. The ions are neutralized by an AN process in which electrons transfer from the sample to the projectiles with the neutralization probability being generally proportional to the amount of time the projectile spends in the near-surface region [12]. AN is an irreversible process in this case because the He ionization level is positioned well below the conduction band of metals and reionization of the projectiles is limited [17]. The spectra thus consist primarily of single scattering events from the outermost atomic layers, since most of the He projectiles that penetrate deeply or experience multiple scattering spend enough time interacting with the sample to be neutralized and are thus not detected by the ESA.

There is, however, a strong matrix effect in He LEIS from graphitic carbon leading to an extremely high neutralization probability for scattered ions, as reported in Refs. [13,18,19]. This effect is very prominent for incoming ions with kinetic energies of 2500 eV or below making it difficult to detect any scattered ions. For example, Mikhailov *et al.* found no C SSP in scattering from a graphitic monolayer on clean Re with 1000 eV $He^+$ ions [18]. According to the calculation carried out by Himpsel, et al. [20], the C 2s and C $2p_z$ states are less bound by about 8 eV in carbidic than in graphitic carbon and the bottom of the sp valence band of graphite is very close to the ground state of Helium 1s (24.6 eV). Thus, the He ions undergo a quasi-resonant neutralization in conjunction with AN in scattering from graphitic carbon [21] leading to an enhancement of the neutralization probability. Therefore, to avoid the matrix effect and provide a detectable signal from C in the graphene layer, 3000 eV $He^+$ projectiles are employed here.

Figure 1 shows ion scattering spectra collected with 3000 eV $He^+$ at a scattering angle of 115° from clean Ru(0001), Gr-covered Ru, and Gr-covered Ru after an $O_2$ exposure, along with an inset that shows the LEED pattern for Gr/Ru(0001). C, O and Ru SSPs can be clearly seen in the various LEIS spectra. Analysis of these SSPs positions and areas as a function of $O_2$ exposure and scattering angle is used to determine how $O_2$ molecules interact with the Gr/Ru(0001) surface. Each SSP rides on a background of multiple scattered projectiles, which is larger on the low energy side of each SSP due largely to reionization [17]. The area of each SSP is computed by integrating the peaks after subtracting the multiple scattering baseline. The baseline is found by fitting the shape of the background surrounding the SSP to a polynomial. The area under the baseline is then subtracted from the total area under the curve to determine the intensity of the SSP itself. When comparing the intensity of SSPs due to different elements, a normalization based on the relative differential cross sections is employed [12,22].

Spectrum (a) in Fig. 1 was collected from a clean Ru(0001) surface prepared using several cycles of IBA. A clear Ru SSP at 2600 eV and a sharp hexagonal 1×1 LEED pattern are observed after the cleaning procedure. The shoulder to the left of the Ru SSP is attributed to multiple scattering. Those ions that experience multiple scattering have a higher probability for neutralization and lose more energy than singly scattered ions, which is why their intensity is low compared to that of the Ru SSP. The absence of any other SSPs confirms that the surface is clean. The observation of a sharp 1x1 LEED pattern confirms that the surface is well-ordered.

Spectrum (b) was collected from a sample covered with a monolayer of graphene. After growth of the graphene layer, the Ru SSP is replaced by the C SSP at 1150 eV, which indicates that the surface is covered by a complete Gr layer. The Ru SSP is reduced for two reasons. First, the relatively high neutral fraction of AN means that He ions colliding with the underlying Ru atoms are more likely to be neutralized. Second, in this geometry, most of the Ru atoms are shadowed by the Gr overlayer which precludes single scattering. The small bump around 2600 eV represents a small amount of scattering from sub-surface Ru as the Gr atoms don't line up exactly with the atoms in the substrate and therefore don't completely shadow them.

The fact that the C atoms in Gr and the Ru substrate atoms don't line up, despite both being single crystals, leads to a Moiré LEED pattern for Gr/Ru(0001), as shown in the inset to Fig. 1. The Moiré pattern is caused by the different lattice parameters of the graphene layer (2.46

Å) and the Ru crystal (2.30 Å) and indicates a weak interaction between the Gr and the Ru substrate and the formation of a supercell [23,24].

Spectrum (c) in Fig. 1 was collected from Gr/Ru(0001) after a 12800 L exposure to $O_2$. The existence of the O SSP at 1418 eV indicates the presence of oxygen. The exposure to $O_2$ does not change the intensity of the C SSP, but the Ru signal becomes almost undetectable because of additional shadowing of Ru by oxygen. Furthermore, the Moiré pattern disappears following $O_2$ exposure, which shows that the oxygen decouples the interaction between Gr and the Ru substrate and breaks up the super-lattice structure, as was observed for $O_2$ and CO reactions with Gr/Ru(0001) [25-27].

The scattering angle is used here as means for locating the oxygen atoms. At the larger scattering angles, such as the 115° angle used in collecting the spectra shown in Fig. 1 and illustrated on the left side of Fig. 2, the trajectories are close to the surface normal so that the incoming ions can penetrate more deeply and single scattering from deeper lying atoms is possible. If a more grazing angle is used, such as the 45° scattering angle trajectory shown on the right side of Fig. 2, the projectiles are better shadowed from reaching below the outermost Gr layer and do not experience single scattering from the second layer. Thus, the single scattering signal from atoms below the outermost atomic layer is very weak at such small scattering angles.

Figure 3 shows $He^+$ LEIS spectra collected from Gr/Ru(0001) after a 12800 L exposure to $O_2$ using different scattering angles. As the scattering angle decreases, the area of the O SSP becomes smaller until it disappears completely at a scattering angle of 45°, where only a C SSP is observed. As mentioned above, for a smaller scattering angle, it is more difficult for the incoming ions to reach the sub-surface target atoms so that only the signal from the outermost Gr layer is detected. Comparing the spectra measured at 45° and 115°, it can be concluded that the oxygen is present below the Gr overlayer and none is adsorbed on top of the Gr, i.e., the oxygen is intercalated between the Gr and the Ru substrate. Note that the increased AN for He projectiles scattered from the second layer O atoms can lead to a reduction of the O SSP intensity relative to that of the C SSP so that the areas of the C and O SSPs cannot be quantitatively compared to each other.

In Fig. 4, the intercalation of oxygen is monitored as a function of $O_2$ exposure using 3 keV $He^+$ ion scattering at a scattering angle of 115°, in which the intensity of the O SSP is indicative of the concentration of oxygen in the second atomic layer. Exposures of high purity $O_2$ are performed at a pressure of $1.5\times10^{-6}$ Torr with the sample held at 600 K. The oxygen SSP first appears after a 1600 L exposure and increases with additional exposure.

Figure 5 shows the areas of the carbon and oxygen SSPs as a function of $O_2$ exposure. The area of the C SSP during the $O_2$ exposures remains nearly constant, which indicates that the Gr layer is not covered by oxygen following reaction with $1.5\times10^{-6}$ Torr of $O_2$ at 600 K. The oxygen SSP area increases until an exposure of about 7000 L, at which point it saturates. This result is more obvious in the O to C ratio, also shown in Fig. 5, which indicates that the amount of oxygen present is initially small, but grows until it maximizes after a 7000 L $O_2$ exposure. Also, note that the Moiré spots in the LEED pattern disappear by the time that the sample is saturated with $O_2$.

To further understand the $Gr/O_2/Ru(0001)$ system, experiments are performed in which oxygen is desorbed by annealing, with the results shown in Fig. 6. The upper spectrum in Fig. 6 was collected from the fully saturated sample prior to annealing. Following annealing at temperatures up to 600 K, the spectra do not change. A small decrease of the oxygen SSP starts to occur at 700 K and it keeps decreasing at higher annealing temperatures. After being heated to 800 K, the oxygen peak is absent, suggesting that the desorption process is complete. Note that the characteristic Moiré LEED pattern recovers along with the complete desorption of oxygen. The inset to Fig. 6 shows how the oxygen and carbon SSP areas change with annealing temperature. One thing to note is that both the C and O SSP areas decrease during the desorption process.

To compare the behavior of intercalated oxygen to that of oxygen adsorbed onto Ru(0001), the bare Ru surface was exposed to $O_2$ and then subjected to a series of anneals to induce desorption, as shown in Fig. 7. It is understood that $O_2$ reacted with a bare metal will adsorb dissociatively [28,29]. The temperature needed to desorb this atomic oxygen from the bare Ru(0001) surface is about 1200 K, which is 500 K higher than for oxygen intercalated between Gr and Ru(0001).

In Fig. 8, the decrease of the C and O SSPs and how their relationship changes during the desorption process is shown as a function of $O_2$ exposure. The Gr/Ru(0001) sample is prepared with various oxygen exposures at 600 K. After cooling to room temperature, the sample is then annealed to 1000 K for 5 min to accomplish a complete desorption of the intercalated oxygen. The decrease in the carbon and oxygen SSP areas are then computed and shown as a function of $O_2$ exposure in the figure. These decreases represent the amount of C and O lost during the desorption process. The amount of oxygen lost represents the entire amount that was intercalated. The data indicate that with sufficient $O_2$ exposure, there is a concurrent loss of C, presumably due to an etching reaction between the oxygen and Gr. With increasing oxygen exposure, the absolute values of the loss of the oxygen and carbon SSP areas both increase. The ratio of lost C to lost O, also shown in the figure, decreases quickly with the magnitude of the $O_2$ exposure.

**IV. Discussion**

Oxygen adsorbed onto bare Ru(0001) starts to desorb around 1200 K, as seen in Fig. 7, which is 500 K higher than the temperature needed to remove oxygen intercalated between graphene and Ru(0001), as shown in Fig. 6. This indicates that the oxygen adsorbed on bare Ru is more strongly bound than the oxygen intercalated between Gr and Ru. This suggests that the intercalated oxygen is molecular, while oxygen adsorbed on bare Ru is atomic, as intercalated molecules would be significantly less strongly bound than chemisorbed atomic oxygen adatoms.

In recent work, the $Gr/O_2/Ru(0001)$ system was studied by XPS [31]. In this experiment, the O 1s level was found to have two components and the Ru 3d level showed a component that indicated bonding of oxygen to Ru. The $O_2$ pressure used in ref. [31] was 0.5 Torr, however, which is much higher than in the present experiment, and the XPS data was collected under ambient conditions, both of which could lead to the formation of Ru-O bonds. In addition, ref. [31] reported that the oxygen is removed at a lower temperature (750 K) than in the present experiment (800 K), but in this measurement the temperature was continuously increased at a slow enough heating rate so that the sample was effectively annealed for a longer time than the 5 min used here, which likely accounts for the lower temperature needed to remove the oxygen. In addition, there could some small differences in the reported temperatures that are related to calibration of the absolute surface temperature. Thus, the XPS data of ref. [31] are not inconsistent with the conclusion that the intercalated oxygen is molecular under the present conditions that involve lower exposures and measurements in UHV.

Another possible contribution to the differences of the thermal stabilities between oxygen intercalated in Gr/Ru(0001) and adsorbed onto bare Ru(1000) is that the confinement effect of the graphene overlayer destabilizes the bond between O and Ru, as discussed in ref. [9]. DFT calculations had demonstrated a similar confinement effect for CO intercalated between Gr/Pt(111) where the annealing temperature needed to remove CO from Pt(111) is 50 K higher than the temperature need to desorb CO intercalated between Gr/Pt(111) [26]. In their calculation, the adsorption energy of CO decreases by 0.4 eV as the distance between the Gr overlayer and the Pt substrate drops from 5.91 Å of fully relaxed graphene to 5.3 Å showing that the distance decreases due to the Gr-substrate interaction. The calculation confirms that the CO adsorption energy on Pt(111) is weakened by the Gr overlayer. It can be inferred that the smaller the nanospace, the weaker the molecular adsorption. Nevertheless, the difference in desorption temperature of 500 K observed here is much greater than 50 K, suggesting that there is something aside from a confinement effect that is primarily responsible for the increased thermal stability of O chemisorbed directly onto Ru. Since CO adsorbs as a molecule on Pt group metals and Ru [25,33], while oxygen adsorbs dissociatively, this is a reasonable interpretation.

Meanwhile, the possibility cannot be ruled out that it is the reaction of carbon in Gr with the intercalated oxygen that causes desorption at a lower temperature, regardless of the form of the intercalated oxygen. In fact, both the C SSP and O SSP areas decrease after being heated to 600 K, which indicates some etching of the graphene overlayer. Figure 8 shows the loss of carbon during the removal of $O_2$ by annealing, which suggests the products of the desorption are not solely $O_2$, but also include carbon-containing molecules such as CO or $CO_2$. In addition, the ratio of lost C to lost O drops with the amount of oxygen intercalation, showing that the etching reaction has a higher rate at low $O_2$ coverages, even though the absolute amount of C removed increases with $O_2$ exposure.

It is significant to realize that graphene is typically considered as an inert 2D material that can be used as a protection layer. This was investigated here by exposing an as-prepared Gr/Ru sample to 10000 L of $O_2$ at room temperature (data not shown). This surface shows no detectable oxygen SSP with LEIS, confirming that Gr acts as protection layer at room temperature. When the temperature is raised, however, small molecules such as $O_2$ do intercalate between a graphene overlayer and substrates such as Ir(111) and Ru(0001) [8,31,34]. Then, further heating can lead to a substantial number of the intercalated oxygen atoms reacting with the carbon atoms thereby etching the Gr.

In addition, the presence of intercalated oxygen molecules decouples the graphene-metal interaction [8,10,26,35]. This decoupling effect is significant for those transition metals on which Gr is weakly bonded through van der Waals forces. This explains the disappearance of the Moiré pattern of Gr/Ru(0001) after oxygen intercalation. Similar phenomena were observed with scanning tunneling microscopy (STM) for Gr/Ru(0001) [35], electron energy loss spectroscopy (EELS) for CO intercalation between Gr and Pt(111) [26] and XPS and LEED for Gr on SiC [36]. For Gr/Pt(111), it was found that the characteristic loss feature (4-7 eV) due to the collective excitation of $\pi$ electrons in freestanding graphene [37] is not observed on the Gr/Pt(111) surface [26]. The absence of this feature is believed to be caused by an electronic interaction between graphene and the Pt substrate that disrupts the $\pi$ band structure, and the presence of intercalated CO causes the loss feature to return [26]. Thus, results from the literature indicate that the graphene decouples from the substrate and behaves more like a freestanding layer in the presence of intercalates.

## V. Conclusions

Helium low energy ion scattering is a very useful tool for the study of molecules intercalated between graphene and a substrate. It has already been reported that small molecules, such as $O_2$, intercalate between graphene and a substrate using XPS and microscopy techniques [8-11]. The work presented here, however, unambiguously demonstrates that oxygen is intercalated and not adsorbed, as it is based on a physical technique that directly measures the location of the oxygen, rather than relying on chemical information or images to infer that location. It is further shown that the intercalated oxygen fully desorbs from the surface when the sample is annealed to 800 K, while it takes 1200 K to remove oxygen from bare Ru. It is thus concluded from this large difference in the stability of the reacted oxygen that the intercalated oxygen is molecular. In addition, some of the graphene is etched along with the thermal desorption of oxygen. The products of this desorption likely include $O_2$ along with CO and/or $CO_2$, and the ratio of the products depends on the amount of intercalated oxygen. This implies that special attention is required when heating graphene-based devices in air or low vacuum, where a significant amount of oxygen is present, to avoid etching that could result in degradation of the quality of the graphene material. Also, the property that the graphene can ease the desorption of adsorbates from certain substrates might have a potential use in catalysis.

## VI. Acknowledgments

This material is based upon work supported by the National Science Foundation under CHE - 1611563.

30. K. S. Kim and N. Winograd, J. Catal. **35**, 66 (1974).

31. A. Dong, Q. Fu, M. Wei, Y. Liu, Y. Ning, F. Yang, H. Bluhm, and X. Bao, Surf. Sci. **634**, 37 (2015).

32. K. Siegbahn, *ESCA applied to free molecules* (North-Holland Pub. Co., 1970).

33. F. Gao, Y. Wang, Y. Cai, and D. W. Goodman, J. Phys. Chem. C **113**, 174 (2009).

34. E. Grånäs, J. Knudsen, U. A. Schröder, T. Gerber, C. Busse, M. A. Arman, K. Schulte, J. N. Andersen, and T. Michely, ACS Nano **6**, 9951 (2012).

35. E. Voloshina, N. Berdunov, and Y. Dedkov, Scientific Reports **6**, 20285 (2016).

36. S. Oida, F. R. McFeely, J. B. Hannon, R. M. Tromp, M. Copel, Z. Chen, Y. Sun, D. B. Farmer, and J. Yurkas, Phys. Rev. B **82**, 041411 (2010).

37. T. Eberlein, U. Bangert, R. R. Nair, R. Jones, M. Gass, A. L. Bleloch, K. S. Novoselov, A. Geim, and P. R. Briddon, Phys. Rev. B **77**, 233406 (2008).

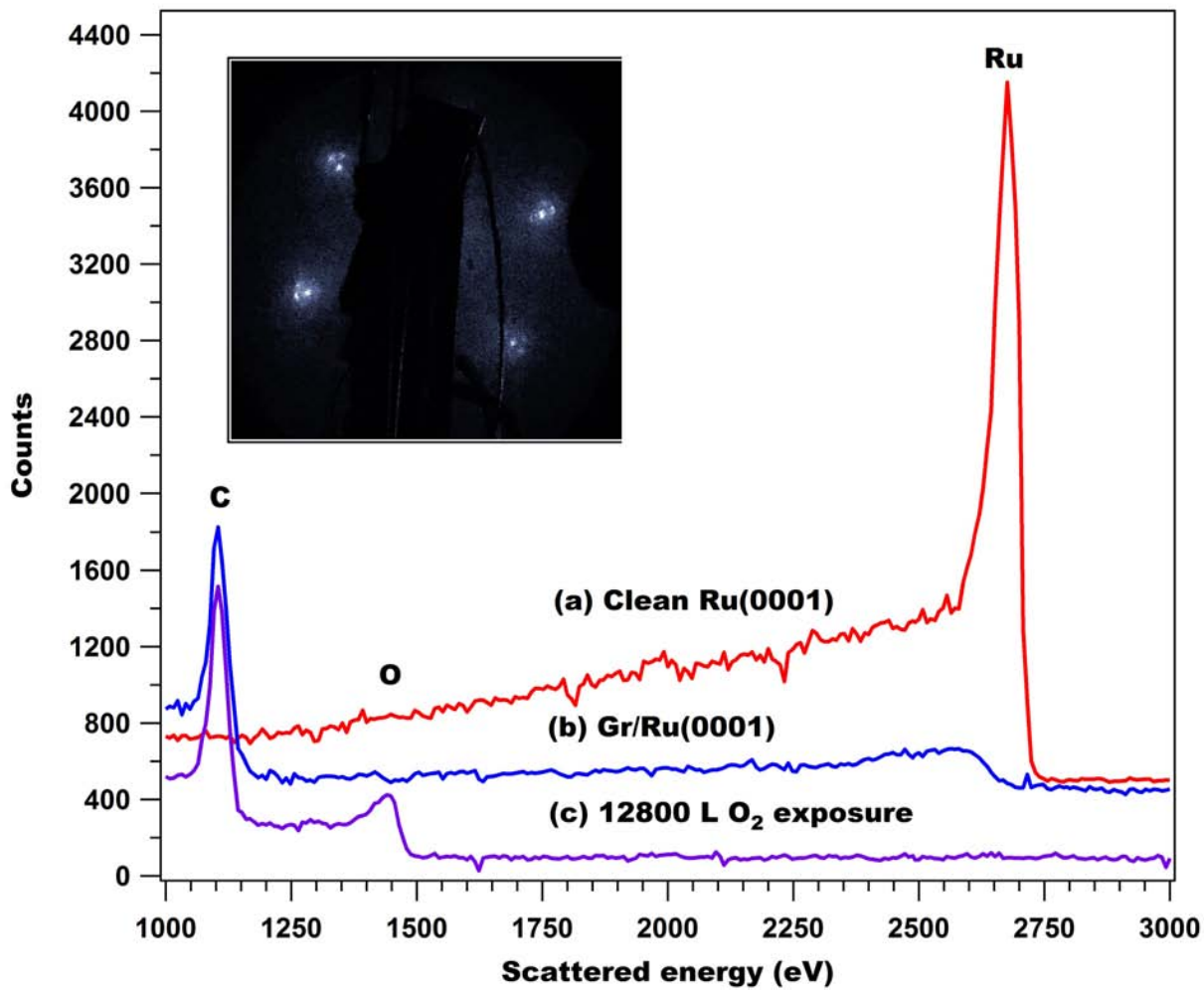

**Figure 1**. 3.0 keV He$^+$ ion scattering spectra collected at a 115° scattering angle from (a) clean Ru(0001), (b) Gr/Ru(0001) and (c) Gr/Ru(0001) exposed at 600 K to 12800 L O$_2$. The carbon, oxygen and Ru SSP positions are labeled. The y-axes are offset for clarity. The inset is the LEED pattern collected from Gr-covered Ru(0001) using an electron energy of 74 eV.

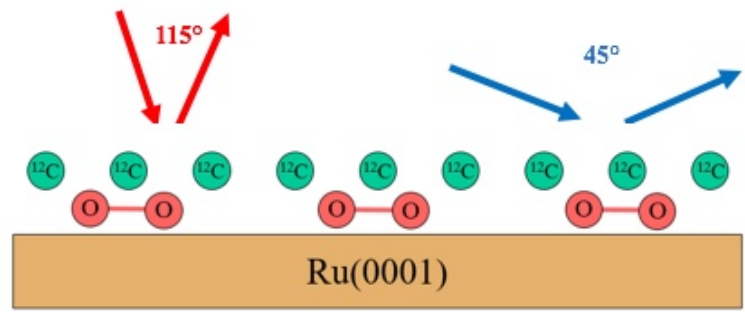

**Figure 2.** A schematic diagram of helium ions scattered from oxygen intercalated Gr/Ru(0001) to show how larger scattering angles can interrogate more deeply below the surface.

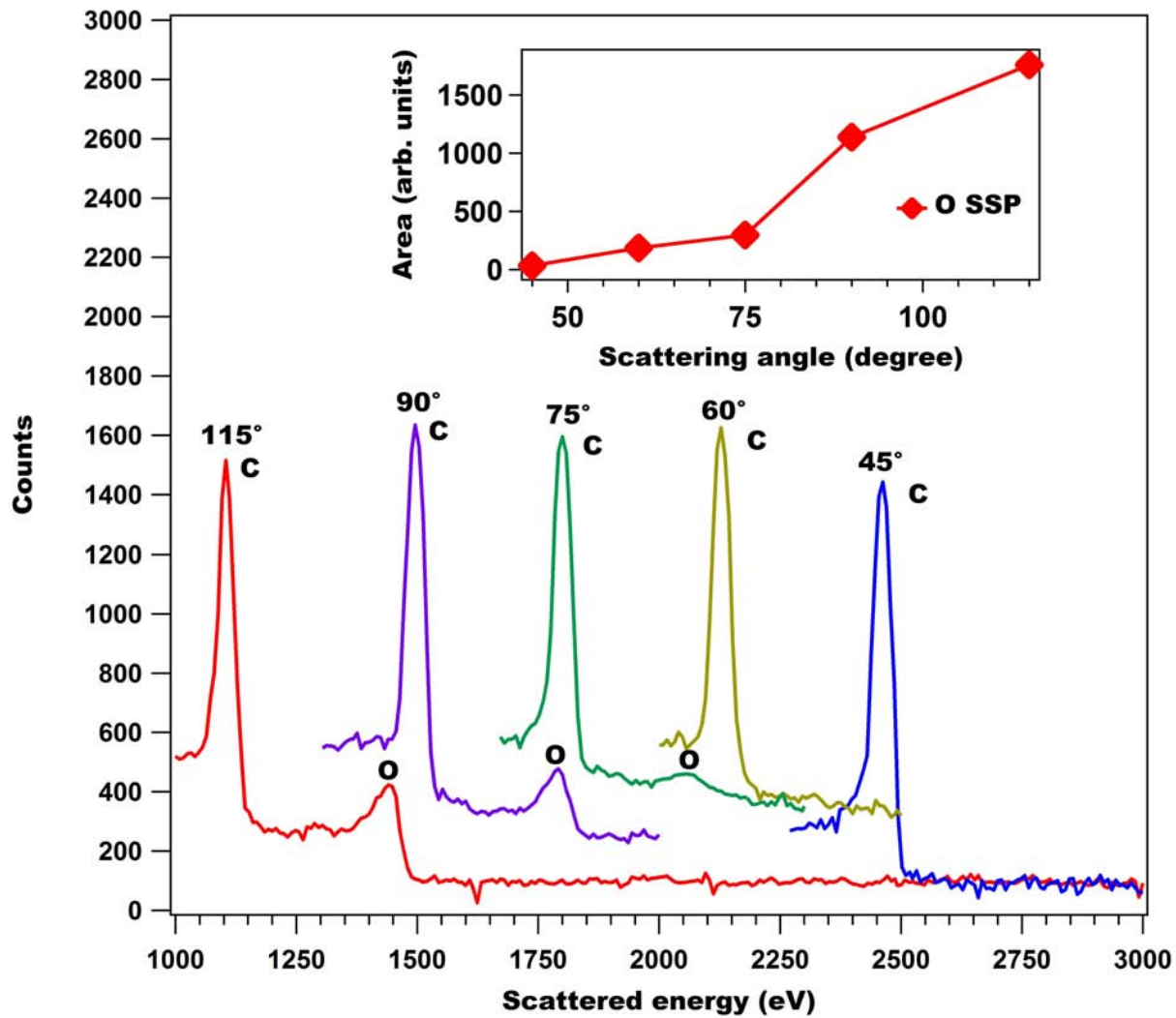

**Figure 3**. 3.0 keV He[+] LEIS spectra, collected from Gr/Ru(0001) exposed at 660 K to 12800 L of $O_2$, at the indicated scattering angles in a specular configuration. The spectra are offset vertically from each other for clarity. The inset shows the O SSP area as a function of the scattering angle.

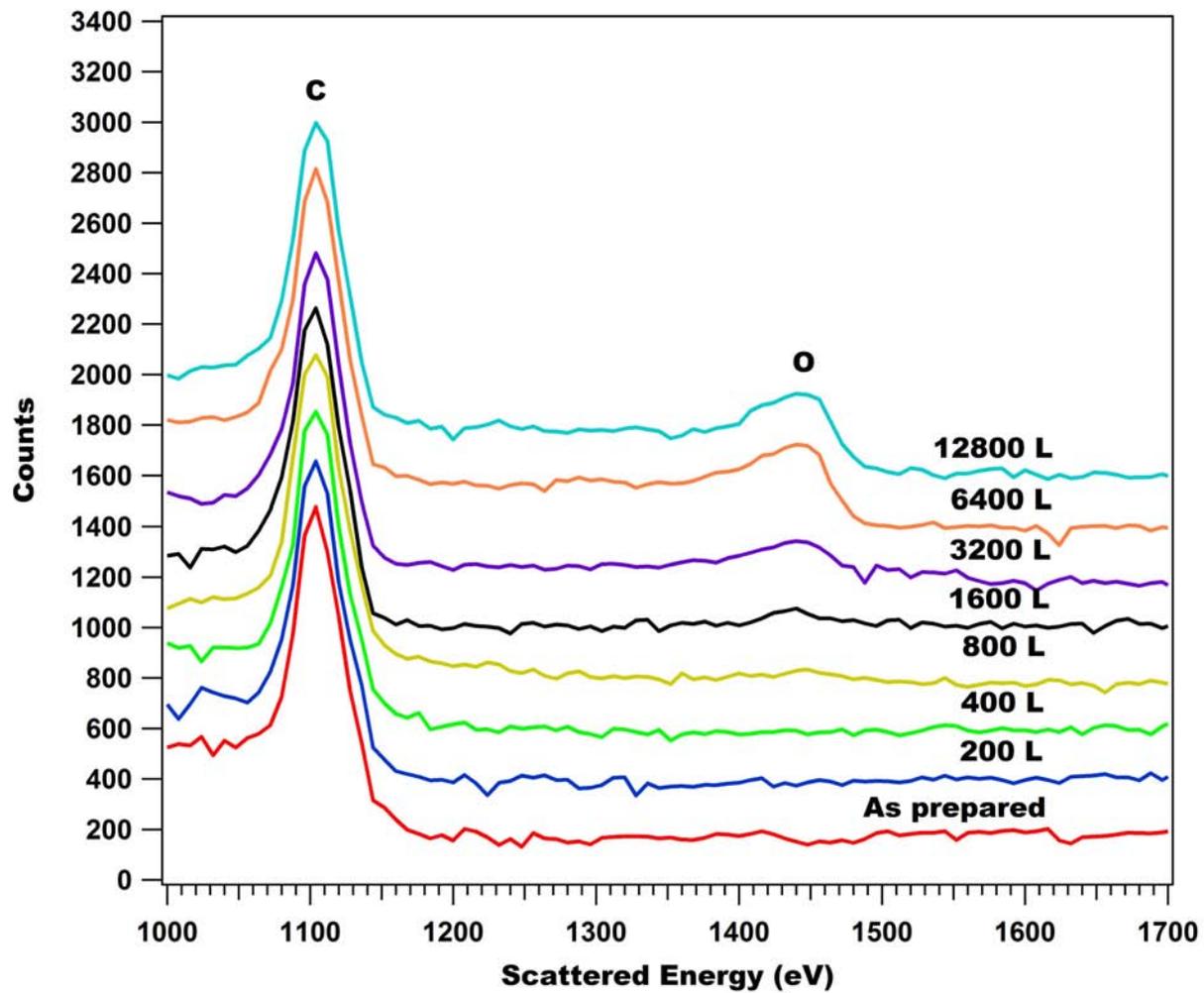

**Figure 4.** 3.0 keV He[+] LEIS spectra collected at a scattering angle of 115° from Gr/Ru(0001) with the indicated $O_2$ exposures.

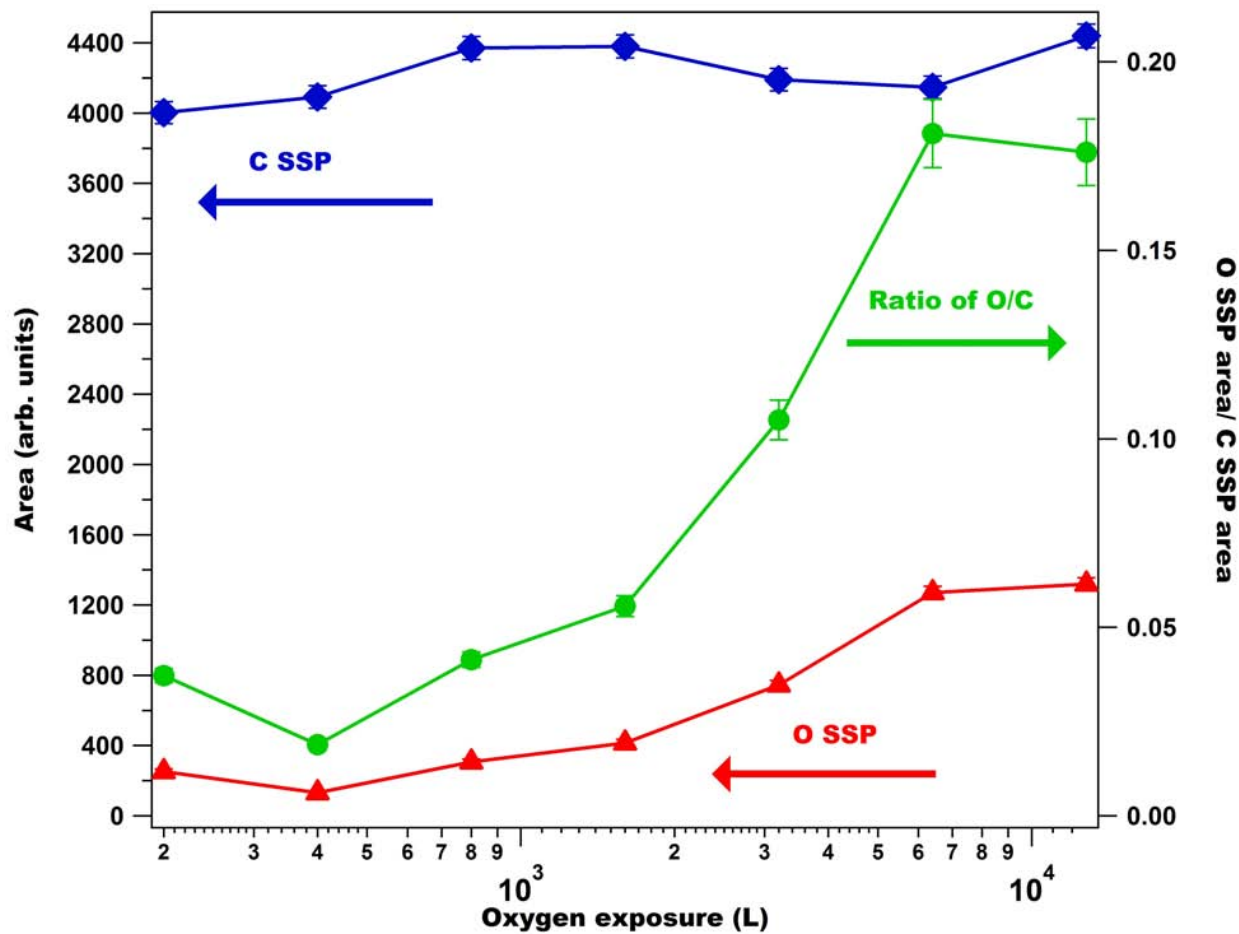

**Figure 5.** The C SSP, O SSP areas and their ratio for $O_2$-exposed Gr/Ru(0001) shown as a function of $O_2$ exposure on a log scale. The ratio curve was produced by dividing and SSP areas and then normalizing by their respective differential scattering cross sections.

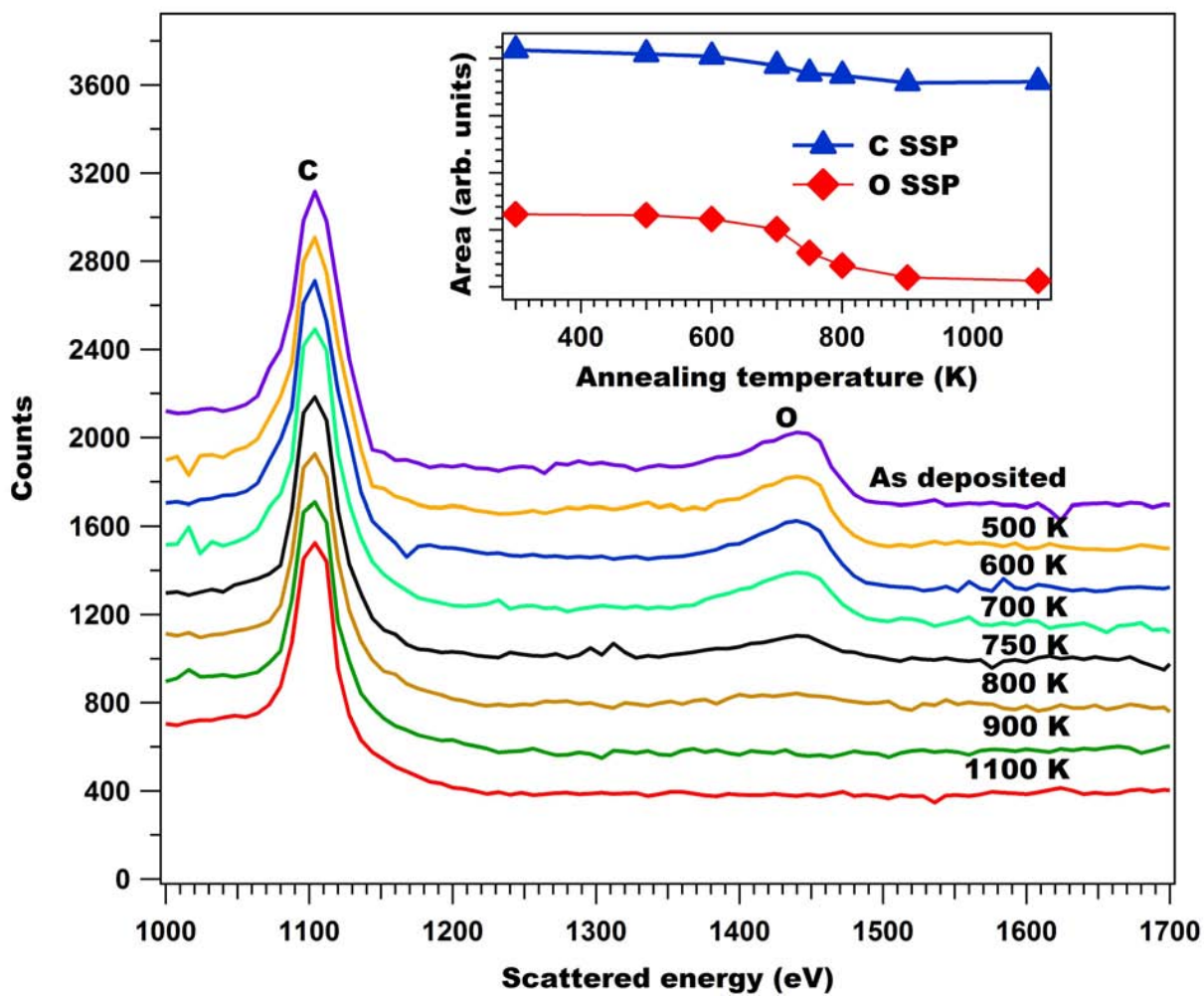

**Figure 6.** LEIS spectra of 3.0 keV He[+] collected at a scattering angle of 115° from Gr/Ru(0001) exposed to 12800 L of $O_2$ at 600 K followed by an additional annealing at the indicated temperature for 5 min. The upper curve in the figure is the as-prepared Gr/$O_2$/Ru(0001). The spectra are offset for better clarity. The inset shows how the C SSP and O SSP areas change with annealing temperature.

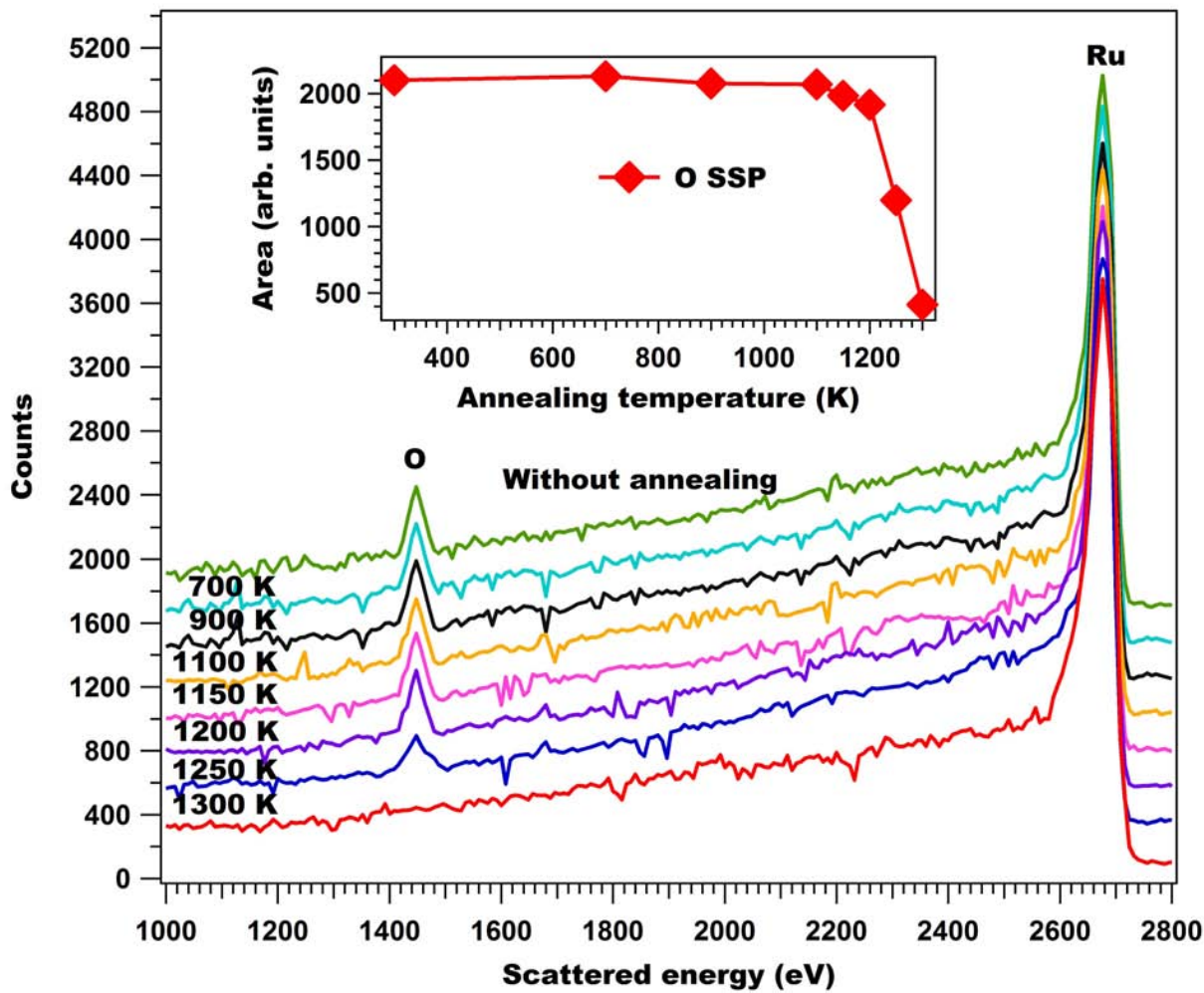

**Figure 7.** LEIS spectra of 3.0 keV He$^+$ collected at a scattering angle of 115° from Ru(0001) exposed to 3000 L of O$_2$ at 400 K, followed by a 5 min annealing at the indicated temperature. The spectra are offset from each other for better clarity. The inset shows how the O SSP area changes with annealing temperature.

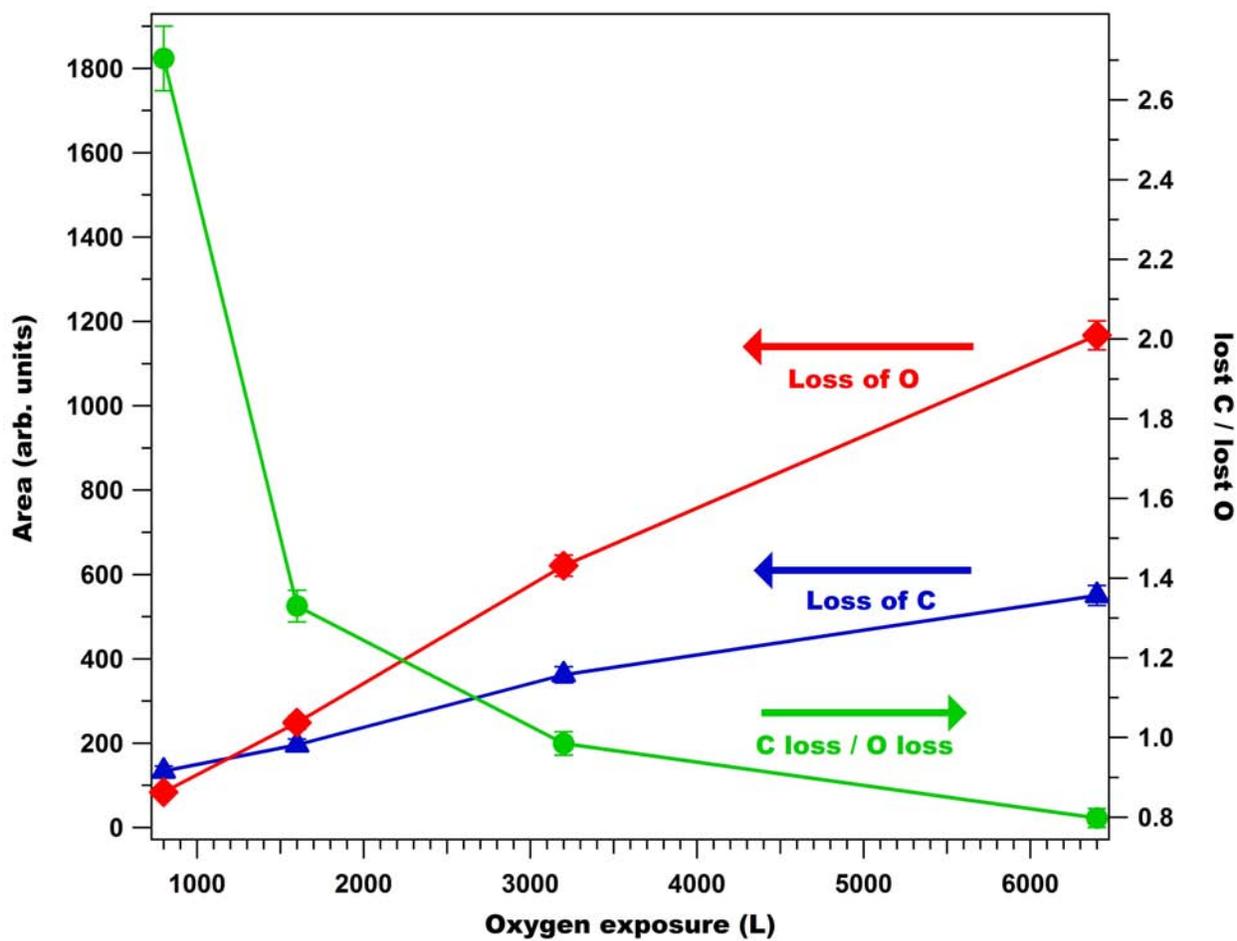

**Figure 8**. The decrease of the oxygen and carbon SSPs during the desorption process as a function of the initial $O_2$ exposure. The Gr/Ru(0001) sample was held at 600 K during the exposures. After cooling, desorption was performed at 1000 K for 5 min. Also shown is the ratio of the decrease of the C to the decrease of the O SSP after normalization by the differential cross sections.